\documentclass[aps,prd,preprint,superscriptaddress,epsf,tightenlines,%
nofootinbib]{revtex4}
\newcommand{\PRE}[1]{{#1}}   

\usepackage{graphicx}
\usepackage{amsmath,amssymb}
\usepackage{bm}
\usepackage{latexsym}

\newcommand{\comment}[1]{}

\def\thalf{\tfrac{1}{2}}
\def\tfour{\tfrac{1}{4}}
\def\ttfour{\tfrac{3}{4}}
\def\tthh{\tfrac{3}{2}}

\def\be{\beta}

\def\comment#1{{}}

\def\tfrac#1#2{{\textstyle\frac{#1}{#2}}}
\def\be{\begin{equation}}
\def\ee{\end{equation}}
\def\bi{\begin{itemize}}
\def\ei{\end{itemize}}
\def\ed{\end{document}}
\def\bfl{\begin{flushleft}}
\def\efl{\end{flushleft}}
\def\bea{\begin{eqnarray}}
\def\eea{\end{eqnarray}}
\def\ket#1{\left|#1\right>}
\def\comp#1#2{\left<#1\vert #2\right>}

\def\amp#1#2#3{\left<#1\left|#2\right|#3\right>}
\def\pa{\vec p_3}
\def\pb{\vec p_4}
\def\la{\lambda_3}
\def\lb{\lambda_4}
\def\slashchar#1{\setbox0=\hbox{$#1$}           
   \dimen0=\wd0                                 
   \setbox1=\hbox{/} \dimen1=\wd1               
   \ifdim\dimen0>\dimen1                        
      \rlap{\hbox to \dimen0{\hfil/\hfil}}      
      #1                                        
   \else                                        
      \rlap{\hbox to \dimen1{\hfil$#1$\hfil}}   
      /                                         
   \fi}
\newif\ifnref

\nreffalse

\input epsf
\def\figin{\epsfcheck\figin}\def\figins{\epsfcheck\figins}
\def\epsfcheck{\ifx\epsfbox\UnDeFiNeD
\message{(NO epsf.tex, FIGURES WILL BE IGNORED)}
\gdef\figin##1{\vskip2in}\gdef\figins##1{\hskip.5in}
\else\message{(FIGURES WILL BE INCLUDED)}%
\gdef\figin##1{##1}\gdef\figins##1{##1}\fi}
\def\DefWarn#1{}
\def\figinsert{\goodbreak\midinsert}
\def\ifig#1#2#3{\DefWarn#1\xdef#1{fig.~\the\figno}
\writedef{#1\leftbracket fig.\noexpand~\the\figno}%
\figinsert\figin{\centerline{#3}}\medskip\centerline{\vbox{\baselineskip12pt
\advance\hsize by -1truein\noindent\footnotefont{\bf Fig.~\the\figno } #2}}
\bigskip\endinsert\global\advance\figno by1}

\def\hat{\widehat}

\begin{document}

\preprint{
\hfil
\begin{minipage}[t]{3in}
\begin{flushright}
\vspace*{.4in}
LMU--ASC 36/08
\end{flushright}
\end{minipage}
}

\title{Decay widths of lowest massive Regge excitations of open strings
\PRE{\vspace*{0.3in}} }

\author{Luis A. Anchordoqui}
\affiliation{Department of Physics,\\
University of Wisconsin-Milwaukee,
 Milwaukee, WI 53201, USA
\PRE{\vspace*{.1in}}
}

\author{Haim Goldberg}
\affiliation{Department of Physics,\\
Northeastern University, Boston, MA 02115, USA
\PRE{\vspace*{.1in}}
}

\author{Tomasz R. Taylor}
\affiliation{Department of Physics,\\
Northeastern University, Boston, MA 02115, USA
\PRE{\vspace*{.1in}}
}
\affiliation{Arnold Sommerfeld Center for Theoretical Physics\\
Ludwig-Maximilians-Universit\"at M\"unchen,
80333 M\"unchen, Germany
\PRE{\vspace{.1in}}
}

\date{June 2008}
\PRE{\vspace*{.5in}}
\begin{abstract}
\vskip 3mm
\noindent With the advent of the LHC there is widespread interest in
the discovery potential for physics beyond the standard model. In
TeV-scale open string theory, the new physics can be manifest in the
excitation and decay of new resonant structures, corresponding to
Regge recurrences of standard model particles.  An essential input for
the prediction of invariant mass spectra of the decay products (which
could serve to identify the resonance as a string excitation) are the
partial and total widths of the decay products. We present a
parameter-free calculation of these widths for the first Regge
recurrence of the $SU(3)$ gluon octet, of the $U(1)$ gauge boson which
accompanies gluons in D-brane constructions, and of the quark triplet.
\end{abstract}
\maketitle

Particles created by vibrations of relativistic strings populate Regge
trajectories re{\nolinebreak}lating their spins and masses. The
threshold mass $M$ for the production of on-shell string excitations
is related to the Regge slope $\alpha'$:
$M=1/\sqrt{\alpha'}$. Although it is generally expected that $M$ is of
order of the Planck mass, much lower values can be envisaged, even
$M\sim 1$~TeV, provided that spacetime extends into large extra
dimensions~\cite{ant}. The low string mass scenarios require the
presence of open strings stretching between D-branes. In particular,
gluons come from strings with both ends attached to a single stack of
coinciding D-branes while quarks from strings stretching between
branes at angles~\cite{inter}.

In proton collisions at the LHC, Regge states will be produced as soon
as the energies of some partonic subprocesses cross the threshold at
$\hat s>M^2$.  But even below the threshold, virtual Regge excitations
will contribute.  In either case, the computations of the
corresponding cross sections involve decay rates of massive string
excitations. Actually, many discovery signals are sensitive to decay
widths, especially those involving particles produced at large
transverse momenta~\cite{jets,Anchordoqui:2008ac}. In this work, we
discuss decays of the first excited string level.\footnote{See
Refs.\cite{peskin,iengo} for previous discussions of some decay modes of string resonances.}

In open superstring\footnote{We do not discuss bosonic strings because
they contain unstable (tachyon) excitations.} theory bosons
and fermions originate from the ten-dimensional Neveu-Schwarz (NS) and
Ramond~\cite{joe} sectors, respectively. The massless level of
open strings associated to a single stack of $N$ D-branes contains
gauge bosons, gauginos and scalars in the adjoint representation of $U(N)$.
The numbers of gauginos and scalars depend on the dimensionality of D-branes and details of
compactification from ten to four dimensions, specifically on the extent to which supersymmetry
is preserved in four dimensions, {\it i.e}.\ on the number of conserved supercharges.
The string vertex operators creating these particles involve
``internal'' fields of superconformal field theory (SCFT) describing string
propagation on the (model-dependent) compact space.
Throughout this work, we adopt the term ``model-dependent'' to characterize such particles,
in contrast to ``universal'' particles like gauge bosons with the vertices depending
on the SCFT fields describing four spacetime coordinates only
(and their SCFT superpartners). A realistic supersymmetry breaking mechanism
must at the end generate masses for all gauginos and adjoint scalars.

The first excited level of open strings ending on a single stack of D-branes can
contain as many as 128 bosonic and 128 fermionic
degrees of freedom (d.o.f.)  \cite{vanp,Tanii} in the adjoint
representation of $U(N)$. In the maximal case, they form
the full massive spin 2 supermultiplet of ${\cal N}=4$ supersymmetry. However, at the leading order of string
perturbation theory ({\it i.e}.\ disk world-sheets), only a few of these
particles couple, and consequently decay, to four-dimensional gauge
bosons \cite{liu}. In the bosonic (NS) sector, only one
massive spin $J=2$ particle (5 d.o.f.) and one spin $J=0$ particle (1
d.o.f.) can decay into two gauge bosons. These massive bosons are universal, in the
sense explained above. They have model-dependent
fermionic superpartners, which decay into gauginos and gluons.
However, due to supersymmetry breaking, gauginos
acquire masses, therefore such decays are
kinematically suppressed.
There also exist model-dependent $J=1$ massive vector bosons, inaccessible in gluon
scattering, which appear as resonances in some gluino fusion
channels.\footnote{There can be also as many as 41 additional (model-dependent)
scalars. Since they do not couple to gluons, they are excluded from our discussion
and should not be confused with the universal $J=0$ state described below in more detail.
They can also appear in certain gaugino fusion channels.}  The number of these states depends
on supersymmetry: ${\cal N}=1$ requires only one but ${\cal N}=4$ as many as 27 massive vector bosons
to complete a massive spin 2 supermultiplet \cite{zin}. The fate of these particles upon
supersymmetry breaking is unclear;  however they are phenomenologically interesting
because they can appear as resonances in quark-antiquark annihilation at exactly
the same center of mass energies
as the universal $J=2$ and $J=0$.\footnote{We are grateful to Michael Peskin for pointing this out.}

The part of the spectrum originating from strings stretching between branes at angles, including quarks, is model-dependent. Nevertheless, it exhibits some universal features.
Quarks have $J=1/2$ and $J=3/2$ Regge excitations which are accessible in $gq\to gq$ parton scattering. Since the corresponding amplitudes \cite{lhc} do not depend on the compactification details,
the decay rates of these Regge excitations into $qg$ are common to all intersecting brane models.

We will extract the decay rates by factorizing
the polarized four-parton scattering amplitudes on the kinematic poles
(at $\hat s=M^2$) due to Regge excitations propagating in the
intermediate one-particle channels.  This is particularly simple
because at the disk level, all amplitudes with four external gauge bosons
can be extracted
from the so-called MHV amplitude describing just one, maximally
helicity violating configuration $(--+\,+)$
\cite{STii,STi}.\footnote{By convention, all helicities refer to
  incoming particles.} Decay rates into quarks will be extracted in a
similar way.

The $J=0$ particle is a quantum of a completely antisymmetric
three-index tensor field (three-form) $E^{\mu\nu\rho}$ subject to the
constraint $\partial_{\mu}E^{\mu\nu\rho} = 0$, which is dual to a massive
vector field $V_{\mu}$ with a vanishing field strength \cite{taka}.
Its single d.o.f.\ is quite peculiar because it disappears in the
formal $M\to 0$ $(\alpha'\to\infty)$ limit, hence it cannot be
associated to a standard real scalar field.  Due to angular momentum
conservation, this particle decays into two gauge bosons of same helicity,
$(++)$ or $(--)$.  It seems that already at the tree-level, it could
virtually propagate between pairs of gauge bosons and contribute to
``all-plus'' $(+++\,+)$ or ``all-minus'' $(---\,-)$ amplitudes which
are known to vanish.  A possible explanation of this puzzle is that
this single d.o.f.\ is akin to a ``real chiral'' scalar existing in
two non-interacting species coupled to self-dual and anti-self-dual
sectors of Yang-Mills theory, respectively \cite{bill}. One of them is
created by and decays to $(++)$ while the other is created by and
decays to $(--)$. Note that the massive $J=2$ particle is always
created by and decays to two gauge bosons with opposite helicities.

The starting point for computing the decay rates of $J=0$ and $J=2$
states into two gauge bosons is the MHV amplitude
\cite{STii,STi}:\footnote{We use the standard notation of
  \cite{Mangano,Dixon}, although the gauge group generators are
  normalized here in a different way, according to ${\rm
    Tr}(T^{a}T^{b})=\frac{1}{2}\delta^{ab}$.}
\begin{eqnarray}
{\cal
    M}(g^-_1,g^-_2,g^+_3, g^+_4) &=& 4\, g^2\langle
  12\rangle^4 \bigg[  \frac{V_t}{\langle 12\rangle\langle
      23\rangle\langle 34\rangle\langle
  41\rangle}\makebox{Tr}(T^{a_1}T^{a_2}T^{a_3}T^{a_4}+T^{a_2}T^{a_1}T^{a_4}T^{a_3})\nonumber\\ &+&\frac{V_u}{\langle 13\rangle\langle
      34\rangle\langle 42\rangle\langle
21\rangle}\makebox{Tr}(T^{a_2}T^{a_1}T^{a_3}T^{a_4}+T^{a_1}T^{a_2}T^{a_4}T^{a_3})\nonumber\\
      &+&\frac{V_s}{\langle 14\rangle\langle
      42\rangle\langle 23\rangle\langle
31\rangle}\makebox{Tr}(T^{a_1}T^{a_3}T^{a_2}T^{a_4}+T^{a_3}T^{a_1}T^{a_4}T^{a_2})\bigg],
\label{mhv}\end{eqnarray}
where the string ``formfactor'' functions of the Mandelstam variables
$s,t,u~(s+t+u=0)$\footnote{For simplicity, we drop carets for the
  parton subprocesses.} are defined as
\begin{equation}
V_t =V(s,t,u) ~,\qquad V_u=V(t,u,s) ~,\qquad  V_s=V(u,s,t)~,
\end{equation}
with
\begin{equation}
V(s,t,u)= {\Gamma(1-s/M^2)\ \Gamma(1-u/M^2)\over
    \Gamma(1+t/M^2)}\ .
    \end{equation}
    In order to factorize amplitudes on the poles due to lowest
    massive string states, it is sufficient to consider $s=M^2$. In
    this limit, $V_s$ is regular while
\begin{equation}
V_t=\frac{u}{s-M^2}~,\qquad V_u=\frac{t}{s-M^2}~.
\end{equation}
Thus the $s$-channel pole term of the amplitude (\ref{mhv}), relevant
to $(--)$ decays of intermediate states, is
\begin{equation}
{\cal
    M}(g^-_1,g^-_2,g^+_3, g^+_4) \to 4\, g^2\,\makebox{Tr}(\{T^{a_1},T^{a_2}\}\{T^{a_3},T^{a_4}\})\frac{\langle
  12\rangle^4}{\langle 12\rangle\langle
      23\rangle\langle 34\rangle\langle
  41\rangle}\frac{u}{s-M^2}\ .\label{mhvs}
  \end{equation}
The amplitude with the $s$-channel pole relevant to $(+-)$ decays is
\begin{equation}
{\cal
    M}(g^-_1,g^+_2,g^+_3, g^-_4) \to 4\, g^2\,\makebox{Tr}(\{T^{a_1},T^{a_2}\}\{T^{a_3},T^{a_4}\})\frac{\langle
  14\rangle^4}{\langle 12\rangle\langle
      23\rangle\langle 34\rangle\langle
  41\rangle}\frac{u}{s-M^2}\ .\label{mhvo}\end{equation}
  In order to extract decay rates into quark-antiquark pairs, we also need the amplitude \cite{lhc}
  \begin{equation}
{\cal
    M}(q^-_1,\bar q^+_2,g^-_3, g^+_4) =
2\,g^2 {\langle 13\rangle^2\over \langle 14\rangle\langle 24\rangle}\bigg[{t\over s}V_t(T^{a_3}T^{a_4})_{\alpha_1\alpha_2}+{u\over s}V_u(T^{a_4}T^{a_3})_{\alpha_1\alpha_2}\bigg]\ .
\label{quarks}\end{equation}
Its $s$-channel singularity is
\begin{equation}
{\cal
    M}(q^-_1,\bar q^+_2,g^-_3, g^+_4) \to
2\,g^2 \{T^{a_3},T^{a_4}\}_{\alpha_1\alpha_2}\,
\frac{\langle 13\rangle^2}{\langle 14\rangle\langle 24\rangle}\frac{tu}{M^2(s-M^2)}\ .
\label{qlim}\end{equation}

\begin{figure}[tbp]
\begin{center}
\includegraphics[height=7.2cm]{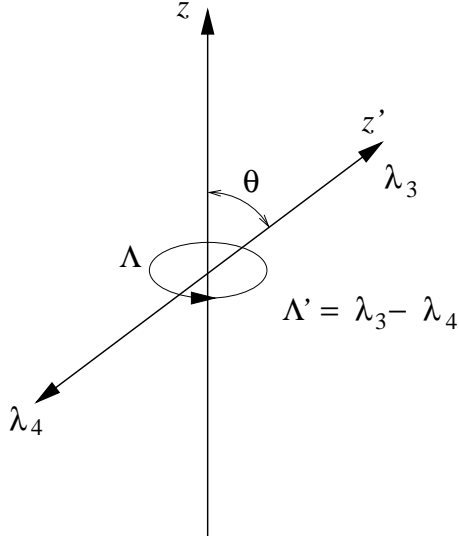}
\caption{Kinematics of the decay of a  spin $J$ resonance, initially
  polarized along the $z$ axis, into two particles whose line of motion
  lies at an angle $\theta$ to the $z$ axis.}
\label{fig:widths}
\end{center}
\end{figure}

In order to use Eqs.~(\ref{mhvs}), (\ref{mhvo}) and (\ref{qlim}), a
softening of the pole term to a Breit-Wigner form, $[s - M^2]^{-1} \to
[(s - M^2) + i \Gamma M]^{-1},$ is necessary.  Here, $\Gamma$ is the
total width of the excitation.  To obtain the partial widths we follow
a standard procedure.  Consider a particle with mass $M$, spin $J$ and
gauge group index $a$ decaying at rest [with momentum $P=(M,\vec 0\,)]$ into particles
$3$ and $4$.  (We will reserve the numbers 1 and 2 for incoming
particles in a scattering situation.)  Let the initial spin component
along the $z$ axis be $J_z =\Lambda.$ The final state may be specified
by the angle $\theta$ of the decay axis $z'$ w.r.t.\ the $z$ axis, the
c.m.\ momenta $\pa,\ \pb= -\pa$, the helicities $\la,\ \lb$ and gauge
indices $a_3,a_4$ of the final state particles. (See
Fig.~\ref{fig:widths}.)

The $S$-matrix element for the decay into particular helicity states
is \be S = i (2\pi)^4\ \delta^4(P-p_3 -p_4)\
\langle\pa\la a_3;\pb\lb a_4|{\cal L}|0, \Lambda,a\rangle \,,
\label{eq:s}
\ee
which in a standard manner gives for the decay width into
these states
\bea
\Gamma_{\la\lb;a_3a_4}^{aJ} &=& \frac{1}{2M}\ \frac{(2\pi)^4}{(2\pi)^6}\
\int\ d^4p_3\ d^4 p_4\ \delta^4(P-p_3 - p_4)
\delta^+(p_3^2-m_3^2)\delta^+(p_4^2-m_4^2) \nonumber \\ [.05in]
&&\hspace*{2in}\times\ \left|\amp{\pa\la a_3;\pb\lb a_4}{{\cal L}}{ 0,
    \Lambda,a}\right|^2
\nonumber \\ [.1in]
&=& \frac{p^*}{32\pi^2\ M^2}\ \int\ d\Omega_3\
\left|\amp{\pa\la a_3;-\pa\lb a_4}{{\cal L}}{ 0, \Lambda,a}\right|^2 \, ,
\label{eq:diffwidth}
\eea where ${\cal L}$ is an interaction Lagrangian, and $p^* = |\pa| =
|\pb| = M/2$ for relativistic particles in the final state.  We now
implement the Wigner expansion into spin eigenstates of the decaying
particle along $z'$~\cite{Yao:2006px}, the decay axis:
\bea
\ket{0,\Lambda}& =& \sum_{\Lambda'}\ket{0,\Lambda'}\comp{0,\Lambda'}{0,\Lambda}\\
&=& \sum_{\Lambda'}\ d^J_{\Lambda\Lambda'}(\theta) \ket{ 0,\Lambda'}\
\ ,
\label{eq:expansion}
\eea
where by rotational invariance $\Lambda' = \la \ - \lb.$ Thus,
\be
\Gamma_{\la\lb;a_3a_4}^{aJ} = \frac{1}{64\pi^2M}\ \int d\Omega \left|F_{\la\lb;a_3a_4}^{aJ}\right|^2\
\left|d^J_{\Lambda,\la-\lb}(\theta)\right|^2 \,,
\label{eq:width2}
\ee where the collinear amplitudes $F^{aJ}$ are matrix elements for the
decay of a particle with $J_{z'} = \la - \lb$ into particles $3,4$
with momenta along the $\pm z'$ axis.
 Since \be \int d\Omega\
\left|d^J_{\Lambda,\ \la-\lb}(\theta)\right|^2 = \frac{4\pi}{2J+1}
\label{eq:norm}
\ee for any $\Lambda$ and helicities, the width of the decaying
particle into a particular helicity pair is \be
\Gamma_{\la\lb;a_3a_4}^{aJ} = \frac{1}{16 (2J+1)\pi M}
\left|F_{\la\lb;a_3a_4}^{aJ}\right|^2\ ,
\label{width3}
\ee independent of the initial spin component $\Lambda$ of the decaying
particle.
Our goal is to extract
$F^{aJ}$
from the center-of-mass resonant scattering amplitude:
\begin{equation}
\langle 3 4; \theta |{\cal M}| 1 2 ; 0\rangle = \sum_{a,J}
\langle34;\theta |{\cal M}^{aJ}| 12;0\rangle\ ,
\end{equation}
where
\begin{equation}
\langle 34;\theta |{\cal M}^{aJ}| 12;0\rangle =  (s-M^2)^{-1}F_{\lambda_3
\lambda_4;a_3a_4}^{aJ} \, F_{\lambda_1 \lambda_2;a_1a_2}^{aJ} \,
d_{\lambda_1-\lambda_2; \lambda_3-\lambda_4}^J (\theta)\ .
 \label{30}
\end{equation}

{}For our purposes, it is convenient to split the $U(N)$ generators
$T^a=(T^0, T^A)$ into $SU(N)$ ``color'' generators $T^A$, $A=1,\dots
N^2-1$ and the $U(1)$ generator $T^0 = \mathbb{I}_N /\sqrt {2 N}$. The
gauge bosons $g^a=(C^0, G^A)$ include the color singlet $C^0$ in
addition to $SU(N)$ gluons $G^A$.  In what follows, we also refer to
the first Regge excitations as $g^{a*} =(C^{0*},G^{A*})$.  The
group-theoretical factors of amplitudes (\ref{mhvs}), (\ref{mhvo}),
and (\ref{qlim}) contain anticommutators
\begin{equation}\{T^{a_1},T^{a_2}\}=4\sum_a d^{a_1a_2a}T^a\ ,\end{equation}
where the completely symmetric $d$ symbol denotes the symmetrized trace
\begin{equation}
d^{a_1a_2a_3}=\makebox{STr}(T^{a_1}T^{a_2}T^{a_3})\ .\end{equation}
Note that, in particular,
\begin{equation}
d^{000}=\frac{1}{\sqrt{8N}}~,\qquad d^{00A}=0~,\qquad d^{0AB}=\frac{1}{\sqrt{8N}}\delta^{AB}~.\end{equation}

We shall first consider the case where all
four particles are $U(N)$ gauge bosons $g^a,\; a= 0,\dots, N^2-1$.
The group-theoretical factor of the resonant amplitudes (\ref{mhvs}) and (\ref{mhvo}) is
\begin{equation}
{\rm Tr} (\{T^{a_1}, T^{a_2}\} \{T^{a_3}, T^{a_4}\}) = 8 \sum_a d^{a_1a_2a} \, d^{a_3a_4a}\equiv Q_{a_1a_2}^{a_3a_4}\ .
\end{equation}
{}From Eqs.\ (\ref{mhvs}) and (\ref{mhvo}) it follows that
\begin{eqnarray}\sum_a
\langle \pm1 \pm 1; \theta |{\cal M}^{aJ=0}| \pm 1 \pm 1, 0 \rangle &=& Q_{a_1a_2}^{a_3a_4}\frac{(2gM)^2}{s-M^2}  \,,
\label{34}\\
\sum_a\langle \pm1 \mp 1; \theta |{\cal M}^{aJ=2}|
\pm 1 \mp 1, 0 \rangle &=& Q_{a_1a_2}^{a_3a_4}\frac{(2gM)^2}{s-M^2}\,  d_{\pm 2,\pm 2}^2 (\theta) \,,
\label{35}\\
\sum_a\langle \mp 1 \pm 1; \theta |{\cal M}^{aJ=2}| \pm 1 \mp 1, 0 \rangle &=&Q_{a_1a_2}^{a_3a_4}
 \frac{(2gM)^2}{s-M^2}\,  d_{\pm 2,\mp 2}^2 (\theta) \ .
\label{36}
\end{eqnarray}

Thus, comparing Eqs.~(\ref{34}), (\ref{35}), and (\ref{36}) with
Eq.~(\ref{30}) we obtain the collinear amplitudes (up to a phase):
\begin{equation}
F_{++a_3a_4}^{aJ=0} =   F_{--a_3a_4}^{aJ=0} = F_{+-a_3a_4}^{aJ=2} = F_{-+a_3a_4}^{aJ=2} = 4\sqrt{2} g\, M  \, d^{a_3a_4a} \, .
\label{37}
\end{equation}
By using Eq.~(\ref{width3}), we obtain the width for the decay of a
resonance $g^{a*}$, with $J=0$ or $J=2$, into two gauge bosons:
\begin{eqnarray}
\Gamma^{J}_{g*\to gg} & = & \frac{1}{2} \times \frac{1}{16(2J+1) \pi M}
\sum_{\lambda_3, \lambda_4} \sum_{a_3,a_4} \,   |F_{\lambda_3 \lambda_4a_3a_4}^{aJ}|^2
\label{38} \\
& = & \left\{ \begin{tabular}{c} $\frac{g^2 M}{(2J+1) \pi}  \, \sum_{a_3,a_4} d^{a_3a_4a} \, d^{a_3a_4a} \ \ \ \ \ \ \ \ \ J=0$ \nonumber\\
$ \frac{2 g^2 M}{(2J+1) \pi}  \, \sum_{a_3,a_4} d^{a_3a_4a} \, d^{a_3a_4a} \ \ \ \ \ \ \ \ \  J=2$  \nonumber
\end{tabular}
\right. ,
\end{eqnarray}
where a factor of $1/2$ have been introduced to take account of either
double counting or identical particles in the final state. Note that
for $J=2$ there
are two possible helicity configurations for the gluons in the final
state, see Eq.~(\ref{37}).  For $J = 0$, the situation is
different because as discussed above, in order to prevent the excitations of
the resonance through a $(--)$ initial state and its subsequent decay
into a $(++)$ final state, one must in effect consider two degenerate
resonances with non-trivial chiral properties, one of which decays
into $(++)$ and the other one into $(--)$.

The color sum in Eq.(\ref{38}) is evaluated for the varying
possibilities of the $SU(N)$ and $U(1)$ assignments:
\begin{itemize}
\item For the decay of  $G^*\to GG,$
\begin{equation}
\sum_{B,C=1}^{N^2-1} d^{ABC} \, d^{ABC} =  \frac{N^2-4}{16N}\ ,
\end{equation}
for any initial $A=1,\dots, N^2-1$~\cite{groupf}.
\item For $G^* \to G C^0,$
\begin{equation}
2\sum_{B = 1}^{N^2-1} d^{AB0} \, d^{AB0} = \frac{1}{4N} \ \
\end{equation}
for any initial $A=1,\ldots N^2-1$.
\item Similarly, for $C^{0*}\to GG$
\begin{equation}
\sum_{B,C = 1}^{N^2-1} d^{BC0} \, d^{BC0} = \frac{N^2-1}{8N}\,\,\ \; ;
\end{equation}
\item and for $C^{0*}\to C^0C^0$,
\begin{equation}
d^{000}\ d^{000} = \frac{1}{8N}\ .
\end{equation}
\end{itemize}

Turning now to fermions, we first consider the process $q_{\alpha_1}
\bar q_{\alpha_2}\to g^{a_3}g^{a_4}$, where $\alpha_1,\alpha_2$ are
group indices in the fundamental and anti-fundamental representations
of $U(N)$, respectively. In this case, the group factor in the
respective amplitude (\ref{qlim}) reads
\begin{equation}
\{T^{a_3}, T^{a_4}\}_{\alpha_1\alpha_2} = 4\sum_a d^{a_3a_4a}\ T^a_{\alpha_1\alpha_2}\equiv Q^{a_3a_4}_{\alpha_1\alpha_2}.
\label{colorf}
\end{equation}
The non-zero resonant amplitudes are
\begin{eqnarray}
\sum_a\langle \pm \thalf \mp \thalf; \theta |{\cal M}^{aJ=2}|
\pm 1 \mp 1, 0 \rangle &=&Q^{a_3a_4}_{\alpha_1\alpha_2} \frac{(gM)^2}{s-M^2}\,  d_{\pm 2,\pm 1}^2 (\theta) \,,
\label{35b}\\
\sum_a\langle \mp \thalf \pm \thalf; \theta |{\cal M}^{aJ=2}| \pm 1 \mp 1, 0 \rangle &=& Q^{a_3a_4}_{\alpha_1\alpha_2}\frac{(gM)^2}{s-M^2}\,  d_{\pm 2,\mp 1}^2 (\theta) \ .
\label{36b}
\end{eqnarray}
The amplitude vanishes for $J=0$.
Since the collinear vertex function for decay of the
resonance $g^{a*}\to g^{a_3} g^{a_4} $ is $4\sqrt{2}gM \, d^{a_3a_4a}$,
we  may identify
the decay vertex $g^{a*}\to q_{\alpha_1} \bar q_{\alpha_2}$ from Eqs.~(\ref{colorf}), (\ref{35b}) and  (\ref{36b}) using factorization:
\begin{equation}
{}F^{aJ=2}_{\pm \thalf \mp \thalf\alpha_1\alpha_2} = \frac{1}{\sqrt{2}}\; T^a_{\alpha_1\alpha_2}\; gM
\label{40}
\end{equation}
{}From Eqs.~(\ref{40}) and (\ref{width3}), the width for decay into
$q \bar q$ is
\begin{eqnarray}
\Gamma_{g*\to q\bar q}& =& \frac{2}{16(2J+1)\pi M}\ \left(\frac{gM}{\sqrt{2}}\right)^2\ \makebox{Tr} (T^aT^a)\\
  &=& \frac{g^2}{160\pi}\  M\quad {\rm per\ flavor,} \qquad J=2\,  .
\label{eq:widthqqbar}
\end{eqnarray}
Note that there are two helicity configurations in the
$q\bar q$ final state. It is worthwhile to note that this value for
the width is the same for all $N^2$ gauge bosons, and is independent
of $N$.

In order to compute the decay rate of $J=1$ gluonic excitation into a quark-antiquark pair, we
proceed in two steps. First, since this resonance is inaccessible through gluon-gluon scattering,
we identify it as an intermediate state in the four-gluino amplitude. To that end, we
use the supersymmetric Ward identity \cite{pt}:\footnote{See Ref.\cite{susy} for a
formal proof of supersymmetric Ward identities in full-fledged superstring theory.}
\begin{equation}
{\cal M}(\Lambda^-_1,\Lambda^+_2,\Lambda^+_3, \Lambda^-_4)~=~ \frac{\langle 23\rangle}{\langle 14\rangle}
{\cal M}(g^-_1,g^+_2,g^+_3, g^-_4) \, ,\label{ward}\end{equation}
which implies that the two amplitudes are equal up to a phase factor. Now the $s$-channel
singularity (\ref{mhvo}), previously attributed to pure $J=2$ propagation, needs
reinterpretation appropriate to $(\pm \thalf \mp \thalf)$ external states:
\begin{equation}
d^2_{+2,+2}(\theta)=\tfour d^2_{+1,+1}(\theta)+\ttfour d^1_{+1,+1}(\theta)\, ,\label{ddis}
\end{equation}
which exhibits both $J=2$ and $J=1$ propagation. By using Eqs.(\ref{37}) and (\ref{ward}),
we obtain the collinear amplitudes
\begin{equation}
{}F^{aJ=1}_{\pm \thalf \mp \thalf a_3a_4} =
4\sqrt{\tthh}\, g\, M  \, d^{a_3a_4a} \, .\label{lamlam}
\end{equation}
In the second step, we consider the the four-fermion amplitude with two gauginos and two quarks. Here again, we use the identity \cite{pt}
\begin{equation}
{\cal M}(q^-_1,\bar q^+_2,\Lambda^-_3, \Lambda^+_4)~=~ \frac{[23]}{[24]}
{\cal M}(q^-_1,\bar q^+_2,g^-_3, g^+_4) \, ,\label{wardq}\end{equation}
and extract the pole term by using Eq.(\ref{qlim}):
\begin{equation}
{\cal
    M}(q^-_1,\bar q^+_2,\Lambda^-_3, \Lambda^+_4) \to
8\,g^2 \sum_a d^{a_3a_4a}\,T^{a}_{\alpha_1\alpha_2}\,
\frac{\langle 13\rangle}{\langle 24\rangle}\,\frac{tu}{M^2(s-M^2)}\ .
\label{qlimm}\end{equation}
In this case, both $J=1$ and $J=2$ propagate in the $s$-channel:
\begin{equation}
tu=\tfour M^2\big[\,d^2_{+1,-1}(\theta) ~+~ d^1_{+1,-1}(\theta)\big]\quad {\rm at}~s=M^2.\label{tudd}
\end{equation}
After extracting the part of (\ref{qlimm}) describing  $q\bar q\to g^*(J=1)\to \Lambda\Lambda$ and factorizing out the  $g^*(J=1)\to \Lambda\Lambda$ vertex  given in Eq.(\ref{lamlam}), we obtain the following collinear amplitude for the decay of the $J=1$ resonance into a quark-antiquark pair:
\begin{equation}
{}F^{aJ=1}_{\pm \thalf \mp \thalf \alpha_1\alpha_2} =
\frac{1}{\sqrt{6}} T^a_{\alpha_1\alpha_2}\, g\, M .    \label{lllam}
\end{equation}
The corresponding decay width is
\begin{equation}
\Gamma_{g*\to q\bar q} = \frac{g^2}{288\pi}\  M\quad {\rm per\ flavor,}\qquad J=1 \,  .
\label{wqqbar}
\end{equation}
It is clear from the use of ${\cal N}=1$ supersymmetry that this particular
$J=1$ state is the vector boson component of the massive spin 2, ${\cal N}=1$ supermultiplet.
As mentioned before, it is a model-dependent particle, hence its properties may be affected
by the supersymmetry breaking mechanism. It is worth mentioning that
Eqs.(\ref{ddis}) and (\ref{tudd}) can be also used to recheck the $J=2$ case (\ref{eq:widthqqbar}).

Finally, we turn to the first Regge recurrences of quarks, color
triplets [or more generally, in the fundamental representation of
$SU(N)$] $q^*_\alpha$ with mass $M$ and spins $J=1/2,~3/2$. They appear as
resonances in quark-gluon scattering. The corresponding amplitudes can
be obtained by crossing from Eq.(\ref{quarks}). The relevant
$s$-channel pole terms are
\begin{eqnarray}
{\cal
    M}(q^-_1,g_2^-,\bar q^+_3,g^+_4) &\to&
2\,g^2 \Big(\sum_{\alpha}T^{a_2}_{\alpha_1\alpha}T^{a_4}_{\alpha\,\alpha_3}\Big)
\frac{\langle 12\rangle^2}{\langle 14\rangle\langle 34\rangle}\frac{u}{M^2(s-M^2)}\quad
{\rm for}~ J=1/2\ ,\label{ggqql}\\
{\cal
    M}(q^-_1,g_2^+,\bar q^+_3,g^-_4) &\to&
2\,g^2 \Big(\sum_{\alpha}T^{a_2}_{\alpha_1\alpha}T^{a_4}_{\alpha\,\alpha_3}\Big)
\frac{\langle 14\rangle^2}{\langle 12\rangle\langle 23\rangle}\frac{u}{M^2(s-M^2)}\quad
{\rm for}~ J=3/2\ .
\label{ggql}\end{eqnarray}
After repeating the same steps as for other resonances, we obtain the following collinear amplitudes:
\begin{equation}
F_{+\thalf+1\alpha_3a_4}^{\alpha J=1/2} =   F_{-\thalf-1\alpha_3a_4}^{\alpha J=1/2} = F_{+\thalf-1\alpha_3a_4}^{\alpha J=3/2} = F_{-\thalf+1\alpha_3a_4}^{\alpha J=3/2} = \sqrt{2}\, g\, M  \, T^{a_4}_{\alpha_3\,\alpha} \, .
\label{colq}
\end{equation}
By using the above result, we obtain the width for the decay of a
quark resonance $q^*_\alpha$, with $J=1/2$ or $J=3/2$, into a quark and gluon:
\begin{eqnarray}
\Gamma^{J}_{q^* \to qg} & = &  \frac{1}{16(2J+1) \pi M}
\sum_{\lambda_3, \lambda_4} \sum_{\alpha_3,a_4} \,   |F_{\lambda_3 \lambda_4\alpha_3a_4}^{aJ}|^2
\label{88} \\
& = & \left\{ \begin{tabular}{c} $\frac{g^2 M}{8(2J+1) \pi}  \, \sum_{\alpha_3,a_4} T^{a_4}_{\alpha\,\alpha_3}T^{a_4}_{\alpha_3\alpha}\ \ \ \ \ \ \ \ \ J=1/2$ \nonumber\\
$ \frac{g^2 M}{4(2J+1) \pi}  \, \sum_{\alpha_3,a_4}T^{a_4}_{\alpha\,\alpha_3}T^{a_4}_{\alpha_3\alpha}  \ \ \ \ \ \ \ \ \  J=3/2$  \nonumber
\end{tabular}
\right. ,
\end{eqnarray}
Note that
for $J=3/2$ there
are two possible helicity configurations in the final
state, see Eq.~(\ref{colq}).  For $J = 1/2$, the situation is
different because similarly to the case of gluonic $J=0$, in order to prevent the excitations of
the resonance through a $(-\thalf -1)$ initial state and its subsequent decay
into a $(+\thalf +1)$ final state, one must in effect consider two degenerate
resonances with non-trivial chiral properties, one of which decays
into $(+\thalf +1)$ and the other one into $(-\thalf -1)$.
The color sum in Eq.(\ref{88}) depends whether the final vector boson is the proper gluon in the adjoint representation of $SU(N)$ or a color singlet:
\begin{itemize}
\item For the decay of  $q^*\to q\,G,$
\begin{equation}
\sum_{a=1}^{N^2-1}\sum_{\beta=1}^N T^{a}_{\alpha\beta}T^{a}_{\beta\alpha}=\frac{N^2-1}{2N}
\ ,
\end{equation}
for any initial $\alpha=1,\dots,N$.
\item Similarly, for   $q^* \to q\,C^0,$
\begin{equation}
\sum_{\beta=1}^N T^{0}_{\alpha\beta}T^{0}_{\beta\alpha}=\frac{1}{2N}
\ .
\end{equation}
\end{itemize}

Our results are summarized in tables~\ref{table1}, \ref{table2},
\ref{table3}, and \ref{table4}.  In table~\ref{table1} partial and
total widths are given for any $N$ and number of flavors $N_f$.  The
entries involving $C^{0*}$ and $C^0$ require some discussion. The
singlet gauge field $C^0_{\mu}$ is in reality a linear combination of
the electroweak gauge boson $Y_\mu$ coupled to hypercharge, and an
orthogonal set of $U(1)$ gauge bosons. Any vector boson $Z'_\mu$,
orthogonal to the hypercharge, must grow a mass $M_{Z'}$ in order to
avoid long range forces between baryons other than gravity and Coulomb
forces.\footnote{The anomalous mass growth allows the survival of
  global baryon number conservation, preventing fast proton
  decay~\cite{Ghilencea:2002da}.}  The widths given in
tables~\ref{table1} and \ref{table2} are premised on the assumption
that corrections of order $(M_{Z'}/M)^2$ are negligible, both in
obtaining matrix elements and in calculating phase space. In
table~\ref{table2} we give numerical values for the widths in the case
of $N=3$ and $N_f =6$. In tables~\ref{table3} and \ref{table4} we give
the corresponding partial and total widths for the lowest Regge excitation
of the quark triplet.\vskip 5mm

\begin{table}[htb]
\vspace*{-0.1in} \caption{Partial and total widths of the lowest
Regge excitation of the $U(N)$ gauge bosons. All quantities are to
be multiplied by $(g^2/4\pi)\; M \simeq 100\ {\rm GeV}\; (M/{\rm
TeV})$.}
\label{table1}
\begin{tabular}{c|@{}|c|c|@{}|c|c|@{}|c|c}
\hline
channel & \multicolumn{2}{@{}c|}{$J=0$} &\multicolumn{2}{@{}c|}{$J=1$} & \multicolumn{2}{@{}c}{$J=2$} \\
\hline
 & ~~$G^*$ & $C^{0*}$ ~~ & ~~$G^*$ & $C^{0*}$~~  & ~~$G^*$ & $C^{0*}$ \\
\cline{2-3} \cline{4-5} \cline{6-7}
     & & & & & & \\
$GG$ & ~~$\frac{N^2 -4}{4N}$~~    &  ~~$\frac{N^2-1}{2N}$~~ & $-$ & $-$ &
~~$\frac{N^2 -4}{10N}$~~ & ~~$\frac{N^2 -1}{5N}$~~  \\
 & & & & & & \\
$GC^0$ & $ \frac{1}{N}  $ & $-$ & $-$ & $-$ & $\frac{2}{5N}$ & $-$ \\
 & & & & & & \\
$C^0C^0$ & $-$ & $\frac{1}{2N} $ & $-$ & $-$ & $-$ & $ \frac{1}{5N}   $\\
 & & & & & &  \\
$q \bar q$ &  0 & 0 & ~~~~$\frac{N_f}{72}$~~~~ & ~~~~$\frac{N_f}{72}$~~~~ & $\frac{N_f}{40}$ &  $ \frac{N_f}{40} $ \\
 & & & & & &  \\
{\rm all} & $\frac{N}{4}$ & $\frac{N}{2}$
& $\frac{N_f}{72}$ & $\frac{N_f}{72}$ &
$\frac{N}{10} + \frac{N_f}{40} $
& $\frac{N}{5} + \frac{N_f}{40}  $ \\
& &  & & & & \\
\hline
\end{tabular}
\end{table}
\newpage
\begin{table}[htb]
\vspace*{-0.1in}
\caption{Partial and total widths, in GeV, of the lowest
Regge excitation of the $U(3)$ gauge bosons. (We have taken $N_f = 6$.)
All quantities are to
be multiplied by $M/{\rm
TeV}$.}
\label{table2}
\begin{tabular}{c|@{}|c|c|@{}|c|c|@{}|c|c}
\hline
channel & \multicolumn{2}{@{}c|}{$J=0$} & \multicolumn{2}{@{}c|}{$J=1$} & \multicolumn{2}{@{}c}{$J=2$} \\
\hline
 & ~~$G^*$ & $C^{0*}$ ~~ & ~~$G^*$ & $C^{0*}$ & ~~$G^*$ & $C^{0*}$ \\
\cline{2-3} \cline{4-5} \cline{6-7}
~~~~$GG$~~~~ & ~~~~$41.6$~~~~ & ~~~~$133.3$~~~~ & ~~~$-$~~~ & ~~~$-$~~~ & ~~~~$16.7$~~~~  & ~~~~$53.3$~~~~ \\
$GC^0$ & $33.3$ & $-$ & $-$ & $-$ & $13.3$ & $-$ \\
$C^0C^0$ & $-$ & $16.7$& $-$ & $-$ & $-$ & $6.6$\\
$q \bar q$ &  0 & 0 & $8.3$ & $8.3$ & $15.0 $ &  $15.0$ \\
{\rm all} & $75.0 $ & $150.0  $ & $8.3$ & $8.3$ & $45.0 $ & $75.0$ \\
\hline
\end{tabular}
\end{table}
\begin{table}[htb]
\vspace*{-0.1in}
\caption{Partial and total widths of the lowest
Regge excitation $q^*$ of the quark in the       
fundamental representation of $SU(N)$. All quantities are to
be multiplied by $(g^2/4\pi)\; M \simeq 100\ {\rm GeV}\; (M/{\rm
TeV})$.}
\label{table3}
\begin{tabular}{c|c|c}
\hline
~~~~~~channel~~~~~~& ~~~~~~$J = 1/2$~~~~~~ & ~~~~~~$J =3/2$~~~~~~ \\
\hline
 & & \\
$q\, G$ & $\frac{N^2 -1}{8\,N}$ & $\frac{N^2 -1}{8\,N}$  \\
& & \\
$q\, C^0$ & $\frac{1}{8\, N}$ & $\frac{1}{8\, N} $\\
& & \\
all & $\frac{N}{8}$ & $\frac{N}{8}$ \\
& & \\
\hline
\end{tabular}
\end{table}
\begin{table}[htb]
\vspace*{-0.1in}
\caption{Partial and total widths, in GeV, of the lowest Regge excitation  
   $q^*$ of the $SU(3)$ quark triplet. (We have taken $N_f = 6$.) 
All quantities are to be multiplied by $M/{\rm TeV}$.}
\label{table4}
\begin{tabular}{c|c|c}
\hline
~~~~~~channel~~~~~~& ~~~~~~$J = 1/2$~~~~~~ & ~~~~~~$J =3/2$~~~~~~ \\
\hline
$q\, G$ & $33.3$ & $33.3$  \\
$q\, C^0$ & $4.2$ & $4.2$\\
all & $37.5$ & $37.5$ \\
\hline
\end{tabular}
\end{table}

In summary, within the framework of TeV-scale open string theory we
have presented model independent partial and total decay widths for
the first Regge excitations of the gluon octet, the accompanying $U(1)$
color singlet, and the quark triplet. This is of most immediate
interest for the LHC, as correct values of the widths are critical in
predicting $\gamma$ + jet and dijet invariant mass spectra.

\section*{Acknowledgments}
We are grateful to Michael Peskin for very useful remarks on the properties of $J=1$ Regge states.
L.A.A.\ is supported by the U.S. National Science Foundation and the
UWM Research Growth Initiative.  H.G.\ is supported by the
U.S. National Science Foundation Grants No PHY-0244507 and PHY-0757959.  The research
of T.R.T.\ is supported by the U.S.  National Science Foundation Grants
PHY-0600304, PHY-0757959 and by the Cluster of Excellence ``Origin and Structure of
the Universe'' in Munich, Germany.  He is grateful to Dieter L\"ust and Stephan Stieberger for
their collaboration and permission to use the results of Ref.\cite{lhc} before
 publication. Any opinions, findings, and
conclusions or recommendations expressed in this material are those of
the authors and do not necessarily reflect the views of the National
Science Foundation.

\ed